\documentclass[twocolumn,prl,aps]{revtex4}
\usepackage{braket,xspace,amsmath,amsfonts,amsthm,amssymb,amsbsy,graphics,color,graphicx,bbm}
\usepackage{epigraph}
\usepackage{soul} 
\setstcolor{red}

\usepackage[breaklinks=true]{hyperref}
\hypersetup{
  colorlinks   = true, 
  urlcolor     = blue, 
  linkcolor    = blue,
  citecolor   = blue 
}
\usepackage{framed}

\newcommand\op[2]{|#1\rangle\!\langle#2|}
\newcommand\ipo[3]{\langle#1 |#2| #3 \rangle}
\newcommand\Tr{\operatorname{Tr}}
\newcommand\ip[2]{\langle#1 |#2 \rangle}
\newcommand\Id{\mathbb{I}}
\newcommand\dg{^\dagger}
\newcommand\var{{\rm Var}}
\newcommand\est{\textbf{Estimate}}
\newcommand\detect{\textbf{Detect}}

\begin{document}
\title{Weak value amplification is suboptimal for estimation and detection}

\author{Christopher Ferrie}
\author{Joshua Combes}

\affiliation{
Center for Quantum Information and Control,
University of New Mexico,
Albuquerque, New Mexico, 87131-0001}

\begin{abstract}
We show using statistically rigorous arguments that the technique of weak value amplification (WVA) does not perform better than standard statistical techniques for the tasks of {  single} parameter estimation and signal detection. Specifically we prove that post-selection, a necessary ingredient for WVA, decreases estimation accuracy and, moreover, arranging for anomalously large weak values is a suboptimal strategy. { In doing so, we explicitly provide the optimal estimator, which in turn allows us to} identify the optimal experimental arrangement to be the one in which all outcomes have equal weak values (all as small as possible) and the initial state of the meter is the maximal eigenvalue of the square of the system observable.  Finally, we give { precise quantitative conditions for when weak measurement (measurements without post-selection or anomalously large weak values) can mitigate the effect of uncharacterized technical noise in estimation}.
\end{abstract}

 \date{\today}

\maketitle


 {\em Weak measurements} (also called gentle or fuzzy measurements), where little information is gained about the system at the benefit of little disturbance to that system, are an old  \cite{Ekstein1971,Barchielli1982,CavesMilburn1987} and well-studied concept \cite{FuchsJacobs2001} that has enabled technologies such as quantum feedback control \cite{WisemanMilburn2010}. The distinct concept of a {\em weak value}, often defined in conjunction with weak measurement, was introduced in 1988 by Aharonov, Albert and Vaidman \cite{Aharonov1988How}.  Weak values are said to have practical uses such as increased sensitivity for the purpose of signal detection or quantum metrology (see an introductory exposition and references in \cite{Dressel2013Understanding}). The technique itself is called \emph{weak value amplification}.

An important distinction must be made between two tasks often taken to be equivalent: (1) increasing detection sensitivity through a shift in the meter position; and (2) increasing the accuracy in estimating a parameter which evokes this shift.  Let us call to these two the tasks of \detect\ and \est, respectively.  These two tasks, being statistical in nature, are much older than quantum theory itself.  The task \detect\ is equivalent to \emph{hypothesis testing} while \est\ is equivalent to \emph{parameter estimation}~\cite{VanTrees}.  We note that weak value amplification (WVA, herein) has been motivated by its potential to improve in \textbf{Estimate} while its practical implementation has  been claimed to aid in the task of \detect\    (see, for example, Hosten and Kwiat \cite{Hosten2008Observation}).  

Previously it has been shown by Knee \emph{et al.} \cite{Knee2013Quantum}, for a particular two qubit ancilla coupled estimation problem, that estimation accuracy using weak values is at best equal to the standard estimation technique (using all the data) and typically worse than the standard technique once the postselection probability is correctly accounted.  Knee \emph{et al.} also show that \emph{decoherence} severely penalizes the weak measurement technique.  Recently Tanaka and Yamamoto \cite{Tanaka2013Does}, in a very general ancilla coupled measurement setting, concluded that weak value amplification is ``useless'' for enhancing estimation accuracy once the postselection probability is correctly accounted in the limit of infinite measurements.  This conclusion is slightly more general than that obtained by Zhu \emph{et al.} \cite{Zhu2011Quantum}, who find the same asymptotic results using the signal to noise ratio.  It has been claimed that the above results are known or expected \cite{Dressel2013Understanding} and the true advantage of WVA is the suppression of technical noise.   However, for particular models of technical noise, Knee and Gauger~\cite{Knee2013Weakvalue} have shown that WVA remains unhelpful.

In this Letter we show that WVA is  { suboptimal} both asymptotically (equivalent to previous results) and for any amount of finite data.  Moreover, whereas previous analyses dealt only with the task \est, here we show that WVA is also not tenable for either task: \est\  or \detect .  That is, there is no sense in which WVA provides an ``amplification'' for  { quantum metrology}~\footnote{ Note that we do not comment on weak values \emph{per se} and whether or not they are useful for some other purpose.}.  These conclusions hold true even in the presence of Gaussian technical noise with an arbitrary correlation function. Moreover, we give the precise conditions, and general sense, under which the ``weakness'' of the measurement can mitigate the technical noise. Finally, in deriving the above results, we provide both the optimal experimental arrangement and estimator for the weak measurement scheme, which does not involve throwing out data or invoke the notion of weak values.  Aside from the obvious practical implications, our results clearly illustrate the distinction between weak \emph{measurement}  and weak \emph{values}.

We begin by examining the prototypical example where weak value amplification has been proposed to yield enormous improvements in estimation of small parameters.  Let us suppose system $B$ is a meter with canonical coordinates $[Q,P]=i$ and the interaction Hamiltonian is $H = O\otimes P$, where $O$ is an observable on system $A$.  The initial wave function $\Phi(q)$ of the meter is a zero-mean Gaussian with variance $\sigma^2$ much larger than the eigenvalue range of $O$.
We assume the interaction parameter is small and expand $U (x) = \exp(-i x H)$ about $x=0$, to first order, to obtain an approximation to the joint probability of obtaining outcome $\ket f$ in a measurement of $O$ and $\ket q$ in a measurement of the meter position. The likelihood function is \cite{Aharonov1988How}
\begin{align}\label{joint}
\Pr(f,q|x) =&\Pr(f)\Pr(q|f,x)\nonumber\\
=& |\ip f i|^2 |\Phi(q - x O_w(f))|^2, 
\end{align}
where 
\begin{equation}\label{eq:wv}
O_w(f) = \frac{\ipo f O i}{\ip f i}
\end{equation}
is the \emph{weak value} (assuming it is real). By postselecting on outcome $f=\checkmark$, an anomalously large shift in the average meter position can be observed. That is $O_w$ can be made large by a clever choice of $\ip \checkmark i\approx 0$. In the weak value amplification literature it is suggested that this large shift can be used to ``amplify'' (read: improve) the sensitivity or efficiency of the statistical tasks \est\ and \detect.

To illustrate our general results, {  we explore an example related to the analysis of Feizpour, Xing and Steinberg}~\cite{Feizpour2011Amplifying} (FXS, herein).  FXS consider additional technical noise on the meter variable ($q_j$) such that the $j$th measured signal is
\begin{equation}\label{noise}
r_j = q_j +\eta_j.
\end{equation}
Where $\eta$ is a noise process characterized by its mean $\langle \eta_j\rangle=0$ and correlation $\langle \eta_j\eta_k\rangle$. It is assumed that there is no initial quantum mechanical noise on the meter such that Eq.~\eqref{noise} simplifies to $r_j = x +\eta_j$.  

Following FXS we will compare the weak value (post-selected) \emph{signal-to-noise ratio} (SNR) to the SNR when the measurement results of system $A$ are ignored. Since the extra noise is zero mean, the ``signal'' is defined as the average shift in the meter position over many measurements
\begin{equation}
\hat x = \frac{1}{N} \sum_{k=1}^N r_j. 
\end{equation}
In statistics, this object is called an \emph{estimator} which we denote by the hat.  Because $\mathbb E[\hat x-x]=0$, the estimator is \emph{unbiased}.  Assuming that $\langle \eta_j \eta_k\rangle = \overline{\eta}^2$, which corresponds to the long correlation time regime, the variance of the estimator is
\begin{equation}
\frac{1}{N^2} \sum_{j,k=1} \langle \eta_j \eta_k\rangle = \overline{\eta}^2. 
\end{equation}
The SNR was defined by FXS as the mean of the estimator (the signal) over its standard deviation
\begin{equation}
\text{SNR} = \frac{x}{\overline{\eta}}.
\end{equation}
This is the SNR of the meter variable ignoring the outcomes of the measurement on system $A$.
Now FXS consider post-selection with success probability $p$ and amplified meter position $O_w$ which can be made arbitrarily large (in theory---although practically there maybe limitations).  The variance in this case remains fixed at $\overline{\eta}^2$ but the average meter position is now $O_w x$.  Thus the signal-to-noise increases to
\begin{equation}
\text{SNR} = \frac{O_w x}{\overline{\eta}}.
\end{equation}
We are supposed to conclude that the SNR can be amplified by an arbitrary amount given by the weak value.

The SNR is intended to be a figure of merit for the purpose of either estimating the value of $x$ (\est) or, a least, detecting its presence (\detect ).  However, both tasks are statistical in nature and, before commenting further on SNR,  we appeal to well-established statistical techniques which are indeed used in most areas of experimental physics.  For the \est\  problem, we use the figure of merit \emph{mean squared error} while for \detect\    we measure performance by the probability of correctly identifying the presence of the interaction.  These are the uncontroversially accepted figures of merit for the problems which WVA is claimed to be beneficial.

We begin with \est.  The following equations and calculations are simplified using a vector notation \footnote{More details of the calculations we present can be found in the section III and IV of the appendix and Ref.~\cite{matrix cookbook}.}.  In particular, since each outcome of the $A$ system measurement, labeled $f$, is associated with its own weak value via Eq. \eqref{eq:wv}, we group those in to a vector of weak values labeled $O_w(\boldsymbol{f})$.  Eq. \eqref{noise} becomes $\boldsymbol{r} = \boldsymbol{q} +\boldsymbol{\eta}$ where $\boldsymbol{\eta}$ is a random variable with a Gaussian (or, ``normal'') distribution with zero mean and covariance matrix $\boldsymbol{K}$.  This is denoted $\boldsymbol{\eta}\sim\mathcal N(0,\boldsymbol{K})$. 

The WVA approach is to take all the data $(\boldsymbol{r},\boldsymbol{f})$ and consider the distribution of the meter variable conditioned on the outcomes of the $A$ system: $\Pr(\boldsymbol{r}|\boldsymbol{f},x)$.  A complete statistical analysis, however, utilizes the joint likelihood function of all data: $\Pr(\boldsymbol{r},\boldsymbol{f}|x)$.  To obtain this, we marginalize over $\boldsymbol{q}$:
\begin{equation}\label{eq:integral}
\Pr(\boldsymbol{r},\boldsymbol{f}|x)= \int \Pr(\boldsymbol{r}|\boldsymbol{q})\Pr(\boldsymbol{q},\boldsymbol{f}|x) d\boldsymbol{q}.
\end{equation}
Via the vector generalization of Eq. \eqref{joint}, we have $\Pr(\boldsymbol{q},\boldsymbol{f}|x) = \Pr(\boldsymbol{f})\Pr(\boldsymbol{q}|\boldsymbol{f},x)$ and both functions left in the integrand are Gaussian; thus the integral itself is also Gaussian.  In vector notation,
\begin{equation}\label{normal}
\Pr(\boldsymbol{r},\boldsymbol{f}|x) \sim \Pr(\boldsymbol{f})\mathcal N(x O_w(\boldsymbol{f}), \boldsymbol{K} + \sigma^2  \boldsymbol{1}), 
\end{equation}
where $\boldsymbol{1}$ is the identity matrix and comes from the original (uncorrelated) statistical noise inherent in the quantum measurement.  

From the well-known Cramer-Rao bound, the best estimator---with the minimum mean squared error---is the maximum likelihood estimator (MLE).  Some matrix calculus leads to
\begin{equation}\label{MLE}
\hat x_{\rm MLE} = \frac{O_w(\boldsymbol{f})^{\rm T}(\boldsymbol{K} + \sigma^2  \boldsymbol{1})^{-1} \boldsymbol{r}}{O_w(\boldsymbol{f})^{\rm T}(\boldsymbol{K} + \sigma^2  \boldsymbol{1})^{-1} O_w(\boldsymbol{f})}.
\end{equation}
The variance in the distribution of this estimator gives the minimum mean squared error performance.  Since $\boldsymbol{r}$ is normally distributed, then the estimator is also normally distributed (since it is a linear transformation of $\boldsymbol{r}$).  Thus
\begin{equation}\label{MLE_dist}
\hat x_{\rm MLE} \sim \mathcal N(x, [O_w(\boldsymbol{f})^{\rm T}(\boldsymbol{K} + \sigma^2  \boldsymbol{1})^{-1} O_w(\boldsymbol{f})]^{-1}).
\end{equation}
The variance of the estimator can be easily read off:
\begin{equation}\label{mlevar}
\var[\hat x_{\rm MLE}] = \frac{1}{ O_w(\boldsymbol{f})^{\rm T}(\boldsymbol{K} + \sigma^2  \boldsymbol{1})^{-1} O_w(\boldsymbol{f})}.
\end{equation}

At this point it is worth discussing the role of $\boldsymbol{K}$.  It could be argued that precise knowledge of the covariance matrix of the technical noise is impractical.  One defense says that a device with serious metrological applications will have well-characterized noise properties \cite{YanGusByl13}.  We can say something more interesting for the present scenario.  Consider the Taylor series expansion of $(\boldsymbol{K} + \sigma^2  \boldsymbol{1})^{-1}$ about ${\sigma^2} \gg 1$:
\begin{equation}
  \frac{1}{\sigma^2}\left (\boldsymbol{1} + \frac{\boldsymbol{K}}{\sigma^2}\right)^{-1} = \frac{1}{\sigma^2} \left (\boldsymbol{1} - \frac{\boldsymbol{K}}{\sigma^2}+ O\left(\frac{1}{\sigma^4}\right) \right ).
\end{equation}
With this, we can expand expand the variance of the ``gold-standard'' MLE in Eq. \eqref{mlevar}:
\begin{equation}\label{mlevar_o2}
\var[\hat x_{\rm MLE}] = \frac{\sigma^2}{\|O_w(\boldsymbol{f})\|^2} +\frac{O_w({\boldsymbol{f}})^{\mathrm{ T}}{\boldsymbol{K}} O_w({\boldsymbol{f}})}{\|O_w({\boldsymbol{f}})\|^4} + {O}\!\left(\frac{1}{\sigma^2}\right).
\end{equation}

Next, we show the ignoring the noise covariance matrix altogether results in an estimator that matches the ``gold-standard'' to order $\sigma^{-2}$.  Taylor expanding the MLE in Eq. \eqref{MLE} to first order results in what we call the \emph{simplified} maximum likelihood estimator (SMLE):
\begin{equation}
\hat x_{\rm SMLE} = \frac{O_w(\boldsymbol{f})^{\rm T} \boldsymbol{r}}{\|O_w(\boldsymbol{f})\|^2}.
\end{equation}
The variance of this estimator is 
\begin{align}
\var(\hat x_{\rm SMLE}) &=  \frac{O_w(\boldsymbol{f})^{\rm T}(\boldsymbol{K} + \sigma^2  \boldsymbol{1})O_w(\boldsymbol{f})}{\|O_w(\boldsymbol{f})\|^4}\\
& =  \frac{\sigma^2}{\|O_w(\boldsymbol{f})\|^2} +\frac{O_w(\boldsymbol{f})^{\rm T}\boldsymbol{K} O_w(\boldsymbol{f})}{\|O_w(\boldsymbol{f})\|^4},
\end{align}
which matches the variance in Eq. \eqref{mlevar_o2} to order $\sigma^{-2}$, as promised.  Let us reiterate: the SMLE estimator does not require knowledge $\boldsymbol{K}$, is unbiased and near optimal provided $\sigma^2\gg\|\boldsymbol{K}\|$.  This is gives a precise and quantitative
meaning to the notion that weak measurement (without post-selection and anomalous weak values) can mitigate the effect of technical noise for large enough $\sigma^2$.

Now the WVA technique amounts to choosing a particular outcome $f = \checkmark$ of the $A$ system and keeping only those results of the meter system whose indices correspond to that outcome: $j\in\checkmark$ means $f_j = \checkmark$.  Then the WVA estimator can be written as
\begin{equation}
\hat x_{\rm WVA} = \frac{\sum_{j\in\checkmark} r_j}{N_\checkmark O_w(\checkmark)},
\end{equation}  
where $N_\checkmark \le N$ is the number of times the outcome $\checkmark$ was observed.
This is an unbiased estimator with variance
\begin{equation}
\var(\hat x_{\rm WVA}) = \frac{\sigma^2}{N_\checkmark O_w(\checkmark)^2} + O(1).
\end{equation}
Now, since clearly a sum of positive terms is greater than the sum of a subset of them,
\begin{equation}\label{post}
 \| O_w(\boldsymbol{f})\|^2= \sum_{j=1}^N O_w(f_j)^2 \geq N_\checkmark O_w(\checkmark)^2,
\end{equation}
and
\begin{equation}\label{est_var}
\var[\hat x_{\rm MLE}] \leq \var[\hat x_{\rm SMLE}]\leq\var[\hat x_{\rm WVA}],
\end{equation}
which means that the WVA estimator has the worst squared error (least informative) among the techniques considered here. {  As Eq.~\eqref{est_var} is an inequality it can be saturated, which was first pointed out by Knee et al. \cite{Knee2013Quantum}. }

Let us return the SNR.  Since each of these techniques result in an unbiased estimator, the ``signal''---defined as the mean of the estimator---is $x$.  To first order we have 
\begin{equation}\label{trueSNR}
{\rm SNR}_{\rm MLE} = \frac{x\|O_w(\boldsymbol{f})\|^2}{\sigma^2}\geq \frac{xN_\checkmark O_w(\checkmark)^2}{\sigma^2} = {\rm SNR}_{\rm WVA}.
\end{equation}
In previous analyses, the post-selected SNR was compared to the case where all data from system $A$ was ignored. Equation~\eqref{trueSNR} shows that we can do even better by considering all data.  That is, we have proven that the post-selection portion of the WVA protocol is generally harmful for estimation.  However, it could be that the weak value (or sum thereof) provides an ``amplification'' since the variance is reduced (and SNR increased) by $\|O_w(\boldsymbol{f})\|^2$.  Next, we show that this is false; that is, even if we take account of all data, arranging for some outcomes to have anomalously large weak values { can only increase the variance of the estimator}.

In the variance of the MLE and the SNR, the term $\|O_w(\boldsymbol{f})\|^2$ is a random variable.  This is because we computed the variance of the estimator with respect to the distribution of $\Pr(\boldsymbol{r}|\boldsymbol{f},x)$.  Now we derive the variance with respect to the joint distribution $\Pr(\boldsymbol{r},\boldsymbol{f}|x)$. 
To do this we make use of the {\em law of total variance}: $\var_{\boldsymbol{r},\boldsymbol{f}|x}[\hat x] = \mathbb E_{\boldsymbol{f}}[\var_{\boldsymbol{r}|\boldsymbol{f},x}[\hat x]]+ \var_{\boldsymbol{f}}[\mathbb E_{\boldsymbol{r}|\boldsymbol{f},x}[\hat x]]$.  Since all estimators considered above are unbiased, the second term is zero.  The first term is non-trivial however as it requires the expectation of a ratio of random variables.  To evaluate this, we use again the Taylor series expansion, this time, about the variable $N \gg 1$.  
Still assuming $\sigma^2\gg \|\boldsymbol{K}\|$, we expand the expectation of only the first term in Eq.~\eqref{mlevar_o2}.  The total variance is then \footnote{For a random variable $X$ and a well behaved $g(X)$ we can approximate $ \mathbb E[g(X)]$ by expanding $g(X)=g( \mathbb E[X]) + (X-  \mathbb E[X])(dg/dX) + (1/2)(X- \mathbb E[X])^2(d^2g/dX^2)+ ...$ thus to second order  $ \mathbb E[g(X)] \approx g( \mathbb E[X]) + (1/2)\var(X)(d^2g/dX^2) $.}
\begin{equation}\label{totalvar}
\var[\hat x_{\rm (S)MLE}] = \frac{\sigma^2}{\mathbb E[\|O_w(\boldsymbol{f})\|^2]} + \frac{{ \sigma^2}\var[\|O_w(\boldsymbol{f})\|^2]}{\mathbb E[\|O_w(\boldsymbol{f})\|^2]^3},
\end{equation}
to order $ O({1}/{N^2})$.
Now, $\|O_w(\boldsymbol{f})\|^2 = \sum_{j=1}^N O_w(f_j)^2$ and straightforward calculation reveals  
\begin{align}
\mathbb E\left[ \|O_w(\boldsymbol{f})\|^2 \right] &= N\ipo {i}{O^2}{i},\\
\var\left[ \|O_w(\boldsymbol{f})\|^2 \right] & = N \sum_{k=1}^d p_k (1-p_k)O_w(f_k)^4,
\end{align}
where $d$ is the number of distinct outcomes which occur with probability $p_k $. 
That is, the expected reduction in variance is independent of weak value.  In fact, we see explicitly that the optimal {estimation} strategy is to choose the initial state of the system $A$ to be the eigenvector of $O^2$ with largest eigenvalue.  Plugging these back into Eq. \eqref{totalvar} and dropping the higher order terms gives
\begin{equation}
\var[\hat x_{(S)MLE}] = \frac{1}{ N\ipo {i}{O^2}{i}} + \frac{\sum_{k=1}^d p_k (1-p_k)O_w(f_k)^4}{ N^2\ipo {i}{O^2}{i}^3}.
\end{equation}
Thus, the lowest variance is obtained by choosing the initial state $|i\rangle$ to maximize $\ipo {i}{O^2}{i}$ and minimizing the second term.  Here is the key point: the second term can be forced to zero by taking any of $p_k=1$.  In the weak measurement case this implies that the final measurement basis contains the state $|f\rangle = |i\rangle$.  This will be the only outcome observed and has the weak value $O_w(f) = \ipo{i}{O}{i}$.  These deliberations show that considering ``anomalously large'' weak values  { strictly increases the variance of even the \emph{optimal} estimator---and is hence detrimental} for the task \est.  Next we show the same conclusions are true for the task \detect.

The task \detect\    is equivalent to the statistical problem of \emph{hypothesis testing}.  Let us consider the ``null hypothesis'' that no interaction is present: $x = 0$.  Standard statistical hypothesis testing would have us compute a ``test-statistic''.  In many cases, the most powerful is the \emph{likelihood ratio} test statistic \cite{Neyman1933Problem}:
\begin{equation}
D = -2 \log\left[\frac{\Pr(\boldsymbol{r},\boldsymbol{f}|x=0)}{\max_x \Pr(\boldsymbol{r},\boldsymbol{f}|x)}\right].
\end{equation}
For brevity, we define $\boldsymbol{Q} = (\boldsymbol{K} + \sigma^2  \boldsymbol{1})^{-1}$, so that 
 $\Pr(\boldsymbol{r}|\boldsymbol{f},x) \sim \mathcal N(x O_w(\boldsymbol{f}), \boldsymbol{Q}^{-1})$ and, in particular, $\Pr(\boldsymbol{r}|\boldsymbol{f},x=0) \sim \mathcal N(0, \boldsymbol{Q}^{-1})$.
Then, the log-likelihood ratio becomes
\begin{equation}
D = \boldsymbol{r}^{\rm T} \boldsymbol{Q} \boldsymbol{r} - (\boldsymbol{r}- xO_w(\boldsymbol{f}))^{\rm T} \boldsymbol{Q}(\boldsymbol{r}- xO_w(\boldsymbol{f})).
\end{equation}

According to Wilks' theorem \cite{Wilks1938LargeSample}, under the null hypothesis the distribution of $D$ is asymptotically $\chi^2_N$: the well-known $\chi$-squared distribution with $N$ degrees of freedom.  This fact, or a direct calculation analogous to one below, yields $\mathbb E_{\boldsymbol{r}|\boldsymbol{f},x=0}[D] = N$.  In words: under the null hypothesis---that is, assuming no interaction is present---the distribution of $D$ is a $\chi$-squared random variable and, in particular, its expected value is $N$.  Note that, in practice, this is all we need from the theory; we simply take the data, compute $D$ and if it is sufficiently larger than $N$, we reject the null hypothesis with some degree of confidence.

But, we can in fact do more by designing experiments which are more powerful in that they given larger values of $D$ when an interaction is present.  To this end, we compute the expected value of $D$ when an interaction \emph{is} present.  A lengthy exercise in matrix algebra reveals the expectation value of the two terms of $D$  give
\begin{align}
&\mathbb E_{\boldsymbol{r}|\boldsymbol{f},x}[\boldsymbol{r}^{\rm T} \boldsymbol{Q} \boldsymbol{r}] = N + x^2 O_w(\boldsymbol{f})^{\rm T} \boldsymbol{Q} O_w(\boldsymbol{f}),\\
&\mathbb E_{\boldsymbol{r}|\boldsymbol{f},x}[(\boldsymbol{r}- xO_w(\boldsymbol{f}))^{\rm T} \boldsymbol{Q}(\boldsymbol{r}- xO_w(\boldsymbol{f}))]  = 0.
\end{align}
Summing these two, we have
\begin{equation}
\mathbb E_{\boldsymbol{r}|\boldsymbol{f},x}[D] = N + x^2 O_w(\boldsymbol{f})^{\rm T} \boldsymbol{Q} O_w(\boldsymbol{f}).
\end{equation}
We have already encountered the term $O_w(\boldsymbol{f})^{\rm T} \boldsymbol{Q} O_w(\boldsymbol{f}) = O_w(\boldsymbol{f})^{\rm T}(\boldsymbol{K} + \sigma^2\boldsymbol{1})^{-1} O_w(\boldsymbol{f})$ above.  The Taylor expansion shows
\begin{equation}\label{condit_statistic}
\mathbb E_{\boldsymbol{r}|\boldsymbol{f},x}[D] = N + \frac{x^2}{\sigma^2} \|O_w(\boldsymbol{f})\|^2 + O\left(\frac1{\sigma^4}\right).
\end{equation}
From Eq. \eqref{post} it is clear that the post-selection present in WVA will only reduce this expectation and hence the test has less statistical power.

 { Analogous to the case for \est, now we show that larger weak values have less statistical power.} Consider using all the data to calculate the test statistic $D$ averaged over the outcomes $\boldsymbol{r},\boldsymbol{f}$. Using the {\em law of total expectation} we have $\mathbb E_{\boldsymbol{r}, \boldsymbol{f}|x}[D] = \mathbb E_{ \boldsymbol{f}}[\mathbb E_{\boldsymbol{r} |\boldsymbol{f}, x}[D]]$, ignoring higher order terms this gives
\begin{equation}
\mathbb E_{\boldsymbol{r}, \boldsymbol{f}|x}[D] = N\left(1  + \frac{x^2 \langle i|O^2|i\rangle}{\sigma^2}\right).
\end{equation}
where we have used \eqref{condit_statistic} and the fact that $\mathbb E_{\boldsymbol{f}}[\|O_w(\boldsymbol{f})\|^2] = N\langle i|O^2|i\rangle$. Thus, if an interaction of strength $x$ is present, the test statistic will on average exceed our expectation of $\mathbb E_{\boldsymbol{r}|x=0}[D] = N$ under the null hypothesis by a factor of $1  + x^2 \langle i|O^2|i\rangle / \sigma^2$. Again we see that the weak value amplification technique is suboptimal; the optimal experiment is to have an input state $\ket i$ which maximizes the expected value of $\|O_w(\boldsymbol{f})\|^2$, which occurs when $\ket i$ is eigenvector corresponding to the largest eigenvalue of $O$.  As in the case of \est, the optimal approach to \detect features no anomalously large weak values.


In summary, we have found that post-selection in general can only hinder the ability to perform either the task of \detect\    or \est.  This implies the standard technique of weak value amplification can not provide an improvement for { quantum metrology}.  Moreover, even if all data are processed, it is more advantageous to choose the experimental parameters such that the corresponding weak values are small.  In other words, the typical approach of using ``anomalously large'' weak values is less preferable.  { These negative results are counterbalanced by the following positive ones.} We have identified the optimal input and output states for the system measurement; it is best to choose the input state and output state $\ket{i}$ to be the eigenvector of $O^2$ with maximal eigenvalue.  { We have shown that technical noise can overcomes using weak measurements (without post-selection or anomalous weak values) by choosing a sufficiently broad meter wavefunction.}   

Although the results have been presented by way of an example, they are, as we show in the Supplementary Material~\cite{FerrieCombesSupp}, in fact fully general: post-selection cannot aid in \detect\    or \est\  for any interaction parameter.  Recently, we have generalized the result further to \emph{any} single parameter quantum metrology problem~\cite{ComFerJiaCav13} regardless of how it is imparted on the system.

\begin{acknowledgements}
For helpful correspondences, we thank Carl Caves, George Knee, Aephraim Steinberg, Howard Wiseman, { and ``Referee C''}.  This work was supported in part by National Science Foundation Grant Nos.~PHY-1212445 and PHY-1005540, by Office of Naval Research Grant No.~N00014-11-1-0082 and by the Canadian Government through the NSERC PDF program.
\end{acknowledgements}


\newpage
\onecolumngrid
\appendix
\section{\large Supplementary Material}
\section{General results on estimation}

\subsection{Asymptotic regime Fisher information inequality}

We follow the same approach of Tanaka and Yamamoto \cite{Tanaka2013Does} since it is independent of the strength of the interaction and, being more general, will imply the same conclusion for the weak value amplification regime.  To this end, we compare the \emph{quantum} Fisher information $I_\rho(x)$ of the parameter given the quantum mechanical state before and after postselection.  The quantum Fisher information is defined as $I_\rho(x) = \Tr[\rho(x) L(x)^2]$, where $L$ is implicitly defined through
\begin{equation}\label{SLD}
\frac{\partial }{\partial x} \rho(x) = \frac12 \Big(\rho(x)L(x) + L(x)\rho(x)\Big).
\end{equation}
A more intuitive characterization is that the \emph{quantum} Fisher information is the  \emph{classical} Fisher information of the output probability distribution of a measurement, maximized over all measurement strategies \cite{Braunstein1994Statistical}:
\begin{equation}\label{classical to quantum fisher}
I_\rho(x) = \max_{\{E_k\}} I\big (\Tr [\rho(x) E_k]\big ),
\end{equation}
here $\{E_k\}$ is a positive operator valued measure (POVM).  The reason for using the quantum Fisher information is that it leads to the quantum Cramer-Rao lower bound on the mean squared error of any unbiased estimator \cite{Helstrom1976Quantum}.  
 
The general model is as follows.  There are two systems, $A$ and $B$, which begin in the state $\ket i _A\otimes\ket\phi_B$ {(we drop the subscripts $A$ and $B$ henceforth)} and interact via the unitary $U(x) = \exp(-i x H)$, where $x$ is the unknown parameter of interest.  Under the interaction, the system evolves to $\ket{\psi_{AB}(x)} = U(x) \ket i \otimes\ket\phi$
and after measuring system $A$ in the basis $\ket f$ and postselecting on the outcome $f=\checkmark$, we have the postselected state
\begin{equation} \label{post selected}
\ket{\psi_B(x)} = \frac{\ipo{\checkmark}{U(x)}{i}\ket \phi}{\sqrt{p_\checkmark(x)}},
\end{equation}
where $p_\checkmark(x)$ is the probability of obtaining outcome $\ket \checkmark$.

In \cite{Tanaka2013Does} the main results was as follows: (1) while it is possible to have $I_{\psi_B}(x)\geq I_{\psi_{AB}}(x)$; (2) it is always true that (see section \ref{TanakaYama} of the supplemental Material for a derivation)
\begin{equation}\label{tanaka}
p_\checkmark(x) I_{\psi_B}(x)\leq I_{\psi_{AB}}(x).
\end{equation}
Suppose the total number of measurements is $N$.  Then, as $N\to\infty$, the number of postselected outcomes is $N_\checkmark(x) = p_\checkmark(x) N$.  Rightfully, the authors conclude that the derived inequality \eqref{tanaka} implies that postselection does not help in the limit $N\to\infty$.  As note above, the authors concede, however, that postselection may help in the finite data regime. Now we show that the conclusion in the asymptotic regime holds for finite data as well.  Thus we prove the claim that postselection can not aid in parameter estimation for any amount of data.

\subsection{Rederivation of the Tanaka and Yamamoto result}\label{TanakaYama}
Here we derive Eq.~(9) appearing in Ref.~\cite{Tanaka2013Does} which the authors explicitly show leads to their Eq.~(14) (which we have labeled Eq.~\eqref{tanaka} here). Recall: there are two systems, $A$ and $B$, which begin in the state $\ket i _A\otimes\ket\phi_B$ {(we drop the subscripts $A$ and $B$ henceforth)} and interact via the unitary $U(x) = \exp(-i x H)$, where $x$ is the unknown parameter of interest.  Under the interaction, the system evolves to
\begin{equation}
\ket{\psi_{AB}(x)} = U(x) \ket i \otimes\ket\phi.
\end{equation}
{If we measure systems $A$ in the basis $\ket f$ we may define the Kraus operators $M_f(x)=\bra f U(x)\ket i$ which allow us to specify the POVM elements $E_f(x)=M\dg_f(x)M_f(x)$. From the POVM elements the probability for obtaining outcome $f$ can be calculated $p_f(x)=\Tr [E_f \op{\phi}{\phi}]$}. After measuring system $A$ and postselecting on the particular outcome $f=\checkmark$, we have the postselected state
\begin{equation} 
\ket{\psi_B(x)} = \frac{M_\checkmark(x)\ket \phi}{\sqrt{p_\checkmark(x)}}.
\end{equation}

To calculate the Fisher information of a pure state $\ket{\phi_x}$ with respect to the parameter $x$ we may use
\begin{align} 
I\big( \phi_x \big) = 4\left (  \ip{\partial_x\phi_x}{\partial_x\phi_x} - | \ip{\partial_x\phi_x}{\phi_x}|^2  \right),
\end{align}
where we define $\ket{\partial_x\phi_x} = \partial \ket{\phi_x}/\partial_x$. Lets apply this to the states $\ket{\psi_{AB}(x)}$ and $\ket{\psi_B(x)}$. Starting with $\ket{\psi_{AB}(x)}$ we find
\begin{align} 
I\big( \psi_{AB}(x)   \big) &= 4\left (  \ipo{i,\phi}{H^2}{i,\phi} - | \ipo{i,\phi}{H}{i,\phi}|^2  \right).
\end{align}
The calculation of the Fisher information for $\ket{\psi_B(x)}$ is tedious, we begin by using the quotient rule to obtain
\begin{align} 
\frac{\partial}{\partial x}\! \ket{ \psi_{B}(x) } 
&= \frac{\partial}{\partial x} \frac{M_\checkmark(x)\ket \phi}{\sqrt{p_\checkmark(x)}} 
=\frac{\sqrt{p_\checkmark(x)} \partial_x M_\checkmark(x) -  M_\checkmark(x)  \partial_x \sqrt{p_\checkmark(x)}}{p_\checkmark(x)}\ket \phi .
\end{align}
Now lets massage the term $\partial_x \sqrt{p_\checkmark(x)}$:
\begin{align}
\partial_x \sqrt{p_\checkmark(x)}
&= \partial_x \sqrt{\bra \phi M_\checkmark\dg(x) M_\checkmark(x) \ket\phi}
= \frac{  \ipo{\phi}{\{\partial_xM_\checkmark\dg\!(x) \}M_\checkmark\!(x)+M_\checkmark\dg\!(x) \{\partial_xM_\checkmark\!(x)\}}{\phi}    }{2\sqrt{p_\checkmark(x)}}.
\end{align}
Now we have
\begin{align} 
\frac{\partial}{\partial x}\! \ket{ \psi_{B}(x) } 
&= \frac{\partial_x M_\checkmark(x)\ket \phi}{\sqrt{p_\checkmark(x)}}  -  
\frac{  \ipo{\phi}{\{\partial_xM_\checkmark\dg\!(x) \}M_\checkmark\!(x)+M_\checkmark\dg\!(x) \{\partial_xM_\checkmark\!(x)\}}{\phi}    }{2  \big (p_\checkmark(x)\big)^{3/2}} M_\checkmark(x)\ket \phi.
\end{align}
If we define a scalar $\mathcal{M}_x=\ipo{\phi}{\{\partial_xM_\checkmark\dg\!(x) \}M_\checkmark\!(x)+M_\checkmark\dg\!(x) \{\partial_xM_\checkmark\!(x)\}}{\phi} $ the we have
\begin{align} 
\frac{\partial}{\partial x}\! \ket{ \psi_{B}(x) } \!
&= \frac{\partial_x M_\checkmark(x)\ket \phi}{\sqrt{p_\checkmark(x)}}  -\frac{ \mathcal{M}_x M_\checkmark(x)\ket \phi }{2  \big (p_\checkmark(x)\big)^{3/2}}.
\end{align}

Now we are in the position to state both terms in the Fisher information:
\begin{align} 
 \ip{\partial_x\psi_B(x)}{\partial_x\psi_B(x)}=  \frac{\bra \phi \partial_x M_\checkmark\dg(x) \partial_x M_\checkmark(x)\ket \phi}{p_\checkmark(x)}
  -\frac{\mathcal{M}_x }{2  p_\checkmark(x)^2} \left( 
  \bra \phi \partial_x M_\checkmark\dg(x)  M_\checkmark(x)\ket \phi
 \bra \phi  M_\checkmark\dg(x)  \partial_x M_\checkmark(x)\ket \phi \right)
   +\frac{\mathcal{M}_x^2 }{4  p_\checkmark(x)^2},
\end{align} 
and
\begin{align} 
 |\ip{\partial_x\psi_B(x)}{\psi_B(x)}|^2=  \frac{|\bra \phi \partial_x M_\checkmark\dg(x)M_\checkmark(x)\ket \phi|^2}{p_\checkmark(x)^2}
   -\frac{\mathcal{M}_x }{2  p_\checkmark(x)^2} \left( 
  \bra \phi \partial_x M_\checkmark\dg(x)  M_\checkmark(x)\ket \phi
 \bra \phi  M_\checkmark\dg(x)  \partial_x M_\checkmark(x)\ket \phi \right)
  +\frac{\mathcal{M}_x^2 }{4  p_\checkmark(x)^2}.
\end{align} 
Combining these expressions we find
\begin{align}
I_{\psi_B}(x)=  4\left ( \frac{\bra \phi \partial_x M_\checkmark\dg(x) \partial_x M_\checkmark(x)\ket \phi}{p_\checkmark(x)} -
 \frac{|\bra \phi \partial_x M_\checkmark\dg(x)M_\checkmark(x)\ket \phi|^2}{p_\checkmark(x)^2} \right ),
\end{align}
which agrees with Eq.~(9) of Tanaka and Yamamoto. Then Eq.~(14) of Tanaka and Yamamoto (Eq.~\eqref{tanaka} in our text) follows simply see eg. section \ref{localmeas} of the supplementary material.

\subsection{Finite data}

The final measurement either succeeds ($f=\checkmark$) or does not ($f\neq\checkmark$) with fixed probability $p_\checkmark(x)$.  Labeling each trial $n_i$, we are interested in the total number of successes $N_\checkmark(x) = \sum_{i=1}^N n_i$, which is also a random variable.  The average number of successful postselection events is $\mathbb E[N_\checkmark(x)] = p_\checkmark(x) N$.  Applying inequality \eqref{tanaka} we have
\begin{align}\label{finite_N_mean}
\mathbb E[N_\checkmark(x) I_{\psi_B}(x)]  = p_\checkmark(x) N I_{\psi_B}(x) \leq N I_{\psi_{AB}}(x),
\end{align}
which is our first result and says the following: even for finite data, the average information in the postselected state is less than or equal to that of the full state.  This, however, does not rule out the possibility of obtaining more postselection events than average, from where postselection would provide an ``amplification'' of information.  Next, we show this is unfathomably unlikely.

To rule out the possibility of a particular instance of the random variable $N_\checkmark(x)$ attaining a value greater than its mean $\mathbb E[N_\checkmark(x)] = p_\checkmark(x) N$ in a particular trial we turn to the \emph{Chernoff bound}~\cite{chernoff1952,HoeffdingChernoff1963}. The Chernoff bound is an inequality, for sums of bounded random variables, that bounds the probability that a random variable $X$ is greater than its mean $\mu$ by an amount $\delta\ge0$: $\Pr[X\ge(1+\delta)\mu]\le \exp[ -\delta^2\mu/(2+\delta)]$.  Now we consider the random variable $N_\checkmark(x) I_{\psi_B}$ which represents the attainable Fisher information in all $N$ trials when postselecting on outcome $\checkmark$.  
Now we bound the probability that $N_\checkmark(x) I_{\psi_B}$ can exceed $\mathbb E[N I_{\psi_{AB}}(x)]=N I_{\psi_{AB}}(x)$ in a particular run of an experiment
It follows that  
\begin{equation}
\Pr( N_\checkmark(x) I_{\psi_B}(x) > (1+\delta)N I_{\psi_{AB}}(x)) \leq \exp\!\left( -N p_\checkmark(x) \frac{\delta^2}{2+\delta}\right).
\end{equation}
This is our second result which says the probability of the information of postselection being greater than that of the full state is exponentially suppressed.

\subsection{Local measurements}\label{localmeas}

There is still an objection one can make here.  The inverse of the quantum Fisher information gives a lower bound on the achievable mean squared error in estimating a parameter optimized over all detection strategies, as in Eq.~\eqref{classical to quantum fisher}.  Thus, it could be the case that $I_{\psi_{AB}}(x)$ is larger than $I_{\psi_B}(x)$ yet the former is not achievable due to technical limitations---requiring, for example, entangled measurements.  We can, however, derive an inequality which is stronger than the above bound when a separable measurement is made.

Let us consider the case of a measurement being made on system A in the basis $\{\ket f\}$ (of which the postselected state $\ket 
\checkmark$ is an element).  Rather than postselecting, we will retain all outcomes and ask if the information contained in the collection of states is greater than that of the postselected state.  After the measurement is performed, the state becomes
\begin{equation}
\rho(x) = \sum_f p_f(x) \op f f \otimes \op{\psi_{B,f}(x)}{\psi_{B,f}(x)},
\end{equation}
where, generalizing Eq.~\eqref{post selected}, we have
\begin{align}
\ket{\psi_{B,f}(x)} &= \frac{\ipo{f}{U(x)}{i}\ket \phi}{\sqrt{p_f(x)}},\\
p_f(x) & = \ipo{\psi_{AB}(x)}{(\op f f \otimes \mathbbm 1)}{\psi_{AB}(x)}.
\end{align}
To compute the quantum Fisher information of this state, we would be required to find the operator $L$ implicitly defined in Eq.~\eqref{SLD}.  For our purposes, however, it is enough to see that it will act trivially on the $A$ subsystem (see section (\ref{mainresult}) of the supplemental material for a derivation) .  Then we have 
\begin{equation}\label{post select sum}
I_\rho(x) \geq \sum_f p_f(x) I_{\psi_{B,f}}(x).
\end{equation}
In words, the information contained in the post measurement state is at least the weighted sum of the information in each postselected state.  Since each term in this sum is positive, we are doing strictly worse by postselecting a subset of them.  That is, 
\begin{align}
p_\checkmark(x) I_{\psi_B,\checkmark}(x) \le I_\rho(x)\le I_{\psi_{AB}}(x),
\end{align}  
the last inequality follows from the fact that the quantum Fisher information in optimized over all measurements. This is the main result of our paper.

\subsection{Derivation of Main result}\label{mainresult}

Here we will give the calculation that leads to Eq.~(10).  Recall, after the measurement is performed, the state is
\begin{equation}
\rho(x) = \sum_f p_f(x) \op f f \otimes \op{\psi_{B,f}(x)}{\psi_{B,f}(x)},
\end{equation}
with
\begin{align}
\ket{\psi_{B,f}(x)} &= \frac{M_f(x)\ket\phi}{\sqrt{p_f(x)}}.
\end{align}
It is more convenient for the purpose of this calculation to write this state as
\begin{equation}\label{appendix:rho}
\rho(x) = \sum_f  \op f f \otimes \op{\widetilde{\psi}_{B,f}(x)}{\widetilde{\psi}_{B,f}(x)},
\end{equation}
where now $\ket{\widetilde{\psi}_{B,f}(x)} = M_f(x)\ket\phi$ is an unnormalized postselected state.

In the previous section we used a simplified formula for pure states but to compute the quantum Fisher information in general we need to find the operator $L$ implicitly defined in Eq.~(1).  To this end, we take the derivative of the state in Eq.~\eqref{appendix:rho}:

\begin{align}
\frac{\partial}{\partial x} \rho(x) & = \sum_f  \op f f \otimes \frac{\partial }{\partial x}\left(\op{\widetilde{\psi}_{B,f}(x)}{\widetilde{\psi}_{B,f}(x)}\right),\\
 & = \sum_f  \op f f \otimes \left(M_f(x)\op{\phi}{\phi}{\partial_xM^\dag_f(x)}+\partial_xM_f(x)\op{\phi}{\phi}{M^\dag_f(x)}\right).
\end{align}

Comparing this to Eq.~(1), it is both not obvious and not relevant what the exact form of $L(x)$ is.  What is important, however, is that it only acts non-trivially on system $B$.  That is, $L(x) = \sum_f \op ff \otimes L_f(x)$.  Since $L(x)^2 = \sum_f \op ff \otimes L_f(x)^2$, it follows that  
\begin{align}
I_\rho(x) &= \Tr \left[\rho(x) L(x)^2\right],\\
&= \sum_f p_f(x) \Tr\left [  \op{\psi_{B,f}(x)}{\psi_{B,f}(x)} L_f(x)^2 \right].
\end{align}
Now if we include the normalization we obtain $L_f(x) = \frac{\partial}{\partial x}\log p_f(x) + L_{\psi_f}(x)$ where the latter term is the appropriate $L$ function for $\psi_f$.  Noting that $\sum_f  \frac{\partial}{\partial x} p_f(x) =  \frac{\partial}{\partial x} \sum_f p_f(x) =  \frac{\partial}{\partial x} 1 = 0$, the cross terms disappear and we have
\begin{align}
I_\rho(x) &=  \sum_f p_f(x) \Tr\left [  \op{\psi_{B,f}(x)}{\psi_{B,f}(x)} \left(L_{\psi_f}(x) +  \frac{\partial}{\partial x}\log p_f(x)\right)^2  \right],\\
& =  \sum_f p_f(x) \Tr\left [  \op{\psi_{B,f}(x)}{\psi_{B,f}(x)} \left(L_{\psi_f}(x)^2 +2  \frac{\partial}{\partial x}\log p_f(x)L_{\psi_f}(x)+ \left(\frac{\partial}{\partial x}\log p_f(x)\right)^2\right)  \right],\\
&= \sum_f p_f(x) \left(\frac{\partial}{\partial x}\log p_f(x)\right)^2 + \sum_f p_f(x) I_{\psi_{B,f}}(x).
\end{align}
Since the first term is positive (it is the classical Fisher information in the distribution $p_f(x)$), we arrive at Eq.~\eqref{post select sum}.

\section{General results hypothesis testing}

\subsection{Setup}\label{quantumhypo}
Recall there are two systems, $A$ and $B$, which begin in the state $\ket i _A\otimes\ket\phi_B$ and interact via the unitary $U(x) = \exp(-i x H)$, where $x$ is the unknown parameter of interest.  Under the interaction, the system evolves to
\begin{equation}
\ket{\psi_{AB}(x)} = U(x) \ket i_A \otimes\ket\phi_B.
\end{equation}
Now we measure in the product of the $\ket f_A$ and $\ket r_B$ basis (again we drop the subscripts $A$ and $B$ henceforth), that is we measure the set of projectors $\{\op{f}{f}\otimes\op{r}{r}\}$ which resolve the identity $\sum_{f,r} \op{f}{f}\otimes\op{r}{r}=\Id$. The probability for obtaining outcome $(f,r)$ is
\begin{equation}
\Pr(r,f|x)= \bra{\psi_{AB}(x)} \op{f}{f}\otimes\op{r}{r} \ket{\psi_{AB}(x)}.
\end{equation}
If we make $N$ measurements and label the outcome of a particular trial or measurement by a subscript $j$ we find the likelihood function to be
\begin{align} \label{likelihoodN}
\mathcal{L} (x) \equiv \prod_j^N \Pr(r_j,f_j|x).
\end{align}
Now suppose there are $K$ possible combinations of outcomes $f$ and $r$ which we label as $\{r_k, f_k\}$ where $k\in \{1,\ldots, K \}$. The probability of outcomes $\{ r_1,f_1\},\ldots, \{ r_K, f_K\}$ are $\Pr(r_1,f_1|x),\ldots, \Pr(r_K,f_K|x)$. In a particular run with $N$ trial the outcome $\{r_k, f_k\}$ appears $n_k$ times so that $N= \sum_{k=1}^K n_k$. If we rewrite Eq. (\ref{likelihoodN}) in terms of the number of times outcome $\{r_k, f_k\}$ appears we find the likelihood after $N$ measurements is 
\begin{equation} 
\mathcal{L} (x) = \prod_j^N \Pr(r_j,f_j|x) =\prod_{k=1}^K \Pr(r_k,f_k|x)^{n_k} 
\end{equation}
where $n_k$ is a random variable. On average we expect $\mathbb E[n_k]= N \Pr(r_k,f_k|x)$.

\subsection{Hypothesis testing}
Let us consider the ``null hypothesis'' that no interaction is present: $x = 0$.  Standard statistical hypothesis testing would have us compute a ``test-statistic''.  In many cases, the most powerful is the \emph{likelihood ratio} test statistic \cite{Neyman1933Problem}:
\begin{equation}
D = -2 \log\left[\frac{\Pr({r},{f}|x=0)}{\max_x \Pr({r},{f}|x)}\right].
\end{equation}
According to Wilk's theorem \cite{Wilks1938LargeSample}, under the null hypothesis the distribution of $D$ is asymptotically $\chi^2_N$: a $\chi$-squared distribution with $N$ degrees of freedom. In practice this means we take $N$ data and compute the likelihood ratio test statistic $D$, if $D$ exceeds the value of the $\chi^2_N$ distribution with a chosen level of statistical significance (type I error rate) then we reject the null hypothesis, otherwise we say nothing.

The likelihood ratio test statistic for the model presented in section (\ref{quantumhypo}) is
\begin{equation}
D = -2 \log\left[\frac{ \mathcal{L} (0) }{\max_x  \mathcal{L} (x)  }\right]= -2 \log\left[\frac{ \mathcal{L} (0) }{  \mathcal{L} (\hat x_{\rm MLE})  }\right].
\end{equation}
Simple algebra leads to
\begin{equation}
D =  -2 \log\left[  \prod_{k=1}^K \left ( \frac{  \Pr(r_k,f_k|0)   }{  \Pr(r_k,f_k|\hat x_{\rm MLE})  } \right )^{n_k} \right]=  2\sum_{k=1}^K  {n_k} \log\left[   \frac{  \Pr(r_k,f_k|\hat x_{\rm MLE})  }{  \Pr(r_k,f_k|0)   }  \right].
\end{equation}
It is important to note that this is a sum of non negative terms, that is $\log\left[   {  \Pr(r_k,f_k|\hat x_{\rm MLE})  }/{  \Pr(r_k,f_k|0)   }  \right]\ge 0$, because $ \Pr(r_k,f_k|\hat x_{\rm MLE}) \ge \Pr(r_k,f_k|0)$ by definition of $\hat x_{\rm MLE}$.

\subsection{Postselection and hypothesis testing}
Now we can state our result. First we divide up the outcomes into a set of favorable, $\{r_k, f_k\}\in \checkmark$, outcomes and undesirable outcomes $\{r_k, f_k\}\in \times$ (the division could depend only of $f_k$ or $r_k$ or both). The likelihood ratio test statistic becomes
\begin{equation}
D =    2\left \{ \sum_{k\in \checkmark}  {n_k} \log\left[   \frac{  \Pr(r_k,f_k|\hat x_{\rm MLE})  }{  \Pr(r_k,f_k|0)   } \right ] + \sum_{k\in \times}  {n_k} \log\left[   \frac{  \Pr(r_k,f_k|\hat x_{\rm MLE})  }{  \Pr(r_k,f_k|0)   }  \right]       \right \}.
\end{equation}
Clearly the statistical test becomes weaker if the outcomes $\{r_k, f_k\}\in \times$ are discarded, and thus postselection is harmful for hypothesis testing. This is not to say there are not situations where neglecting the undesirable outcomes will make little difference to the power of the statistical test.

\section{FXS estimation results}

\subsection{Derivation of the maximum likelihood estimator}

Most, or all, of the matrix calculation here can be got at via the formulas in ``The Matrix Cookbook'' \cite{Petersen2012Matrix}. Starting with Eq.~(8) 
\begin{align}
\Pr(\boldsymbol{r},\boldsymbol{f}|x)
=& \int \Pr(\boldsymbol{r}|\boldsymbol{q})\Pr(\boldsymbol{q},\boldsymbol{f}|x) d\boldsymbol{q},\\
=&\Pr(\boldsymbol{f}) \int \Pr(\boldsymbol{r}|\boldsymbol{q})\Pr(\boldsymbol{q}|\boldsymbol{f},x) d\boldsymbol{q}.
\end{align}
Both functions left in the integrand are Gaussian, the integral itself is also Gaussian. One way to see this is to use The Matrix Cookbook~\cite{Petersen2012Matrix} expressions in section 8.1.8: $\mathcal N_q(0,\boldsymbol{K})\,.\, \mathcal N_q(xO_w(\boldsymbol f ), \sigma^2 \boldsymbol{I})= c_c\, \mathcal N_q (m_c, \Sigma_c )$ where $c_c = \mathcal N_{q}(xO_w(\boldsymbol f ), \boldsymbol K + \sigma^2 \boldsymbol I)$, $m_c = [ \boldsymbol{K}^{-1} +(\sigma^2\boldsymbol{I})^{-1}]^{-1}xO_w(\boldsymbol{f})(\sigma^2 \boldsymbol{I})^{-1}$, and $\Sigma_c = [ (\sigma^2 \boldsymbol{I})^{-1} +  \boldsymbol{K}^{-1} ]^{-1}$. But we have to remember that we are then doing an integral: $\int d\boldsymbol q \mathcal N_q (m_c, \Sigma_c ) = 1$. This gives Eq.~(16) of the main text: $\Pr(\boldsymbol{r},\boldsymbol{f}|x) \sim \mathcal N_{\boldsymbol r}(x O_w(\boldsymbol{f}), \boldsymbol{K} + \sigma^2  \boldsymbol{1})$.  To maximize this distribution, it is more convenient to look at the logarithm
\begin{equation}
\log\Pr(\boldsymbol{r},\boldsymbol{f}|x) \propto  (\boldsymbol{r}-xO_w(\boldsymbol{f}))^{\rm T}(\boldsymbol{K} + \sigma^2  \boldsymbol{1})^{-1} (\boldsymbol{r}-xO_w(\boldsymbol{f}))=:\mathcal L(x).
\end{equation}
The derivative of the likelihood with respect to $x$ is obtained using the chain rule, we have
\begin{equation}
\mathcal L'(x) = -2 O_w(\boldsymbol{f})^{\rm T}(\boldsymbol{K} + \sigma^2  \boldsymbol{1})^{-1}( \boldsymbol{r} -x O_w(\boldsymbol{f})).
\end{equation}
Setting this equal to zero and solving for $x$ gives
\begin{equation}\label{a:mle}
\hat x_{\rm MLE} = \frac{O_w(\boldsymbol{f})^{\rm T}(\boldsymbol{K} + \sigma^2  \boldsymbol{1})^{-1} \boldsymbol{r}}{O_w(\boldsymbol{f})^{\rm T}(\boldsymbol{K} + \sigma^2  \boldsymbol{1})^{-1} O_w(\boldsymbol{f})},
\end{equation}
which is Eq.~(10) of the main text.  Now we show that this is distribution according the normal distribution given in Eq.~(11) of the main text.  In general, note that if $\boldsymbol X$ is normally distributed as $\mathcal N (\boldsymbol{\mu},\boldsymbol\Sigma)$ and $\boldsymbol T$ is a linear transformation, then $\boldsymbol T \boldsymbol X$ is distribution according to
\begin{equation}
\mathcal N(\boldsymbol T \boldsymbol\mu, \boldsymbol T \boldsymbol \Sigma \boldsymbol T^{\rm T}).
\end{equation}
Here we transforming the normal variable $\boldsymbol r$ to $\hat x = \boldsymbol T \boldsymbol r$ where 
\begin{equation}
\boldsymbol T = \frac{O_w(\boldsymbol{f})^{\rm T}(\boldsymbol{K} + \sigma^2  \boldsymbol{1})^{-1}}{O_w(\boldsymbol{f})^{\rm T}(\boldsymbol{K} + \sigma^2  \boldsymbol{1})^{-1} O_w(\boldsymbol{f})}.
\end{equation}
Thus, the mean transforms to
\begin{equation}
xO_w(\boldsymbol f) \mapsto \frac{O_w(\boldsymbol{f})^{\rm T}(\boldsymbol{K} + \sigma^2  \boldsymbol{1})^{-1} x O_w(\boldsymbol f)}{O_w(\boldsymbol{f})^{\rm T}(\boldsymbol{K} + \sigma^2  \boldsymbol{1})^{-1} O_w(\boldsymbol{f})} = x,
\end{equation}
and the variance transforms to 
\begin{equation}
\boldsymbol{K} + \sigma^2  \boldsymbol{1} \mapsto \frac{O_w(\boldsymbol{f})^{\rm T}(\boldsymbol{K} + \sigma^2  \boldsymbol{1})^{-1}(\boldsymbol{K} + \sigma^2  \boldsymbol{1})(\boldsymbol{K} + \sigma^2  \boldsymbol{1})^{-1}O_w(\boldsymbol{f})}{(O_w(\boldsymbol{f})^{\rm T}(\boldsymbol{K} + \sigma^2  \boldsymbol{1})^{-1} O_w(\boldsymbol{f}))^2} =\frac{1}{O_w(\boldsymbol{f})^{\rm T}(\boldsymbol{K} + \sigma^2  \boldsymbol{1})^{-1} O_w(\boldsymbol{f})}.
\end{equation}
If we suppose  $\sigma^2\gg \|K\|$, a Taylor expansion reveals
\begin{equation}\label{a:var}
{\rm Var}[\hat x_{\rm MLE}] = \frac{1}{O_w(\boldsymbol{f})^{\rm T}(\boldsymbol{K} + \sigma^2  \boldsymbol{1})^{-1} O_w(\boldsymbol{f})} = \frac{\sigma^2}{\|O_w(\boldsymbol{f})\|^2} +\frac{O_w(\boldsymbol{f})^{\rm T}\boldsymbol{K} O_w(\boldsymbol{f})}{\|O_w(\boldsymbol{f})\|^4} + O\left(\frac{1}{\sigma^2}\right).
\end{equation}

Now suppose the spectral properties of the noise, $\boldsymbol{K}$, are unknown.  In general, this is not a problem since determining the spectral properties of additive noise is a trivial exercise.  One simply runs the experiment without a signal and uses as the maximum likelihood estimator of $\boldsymbol{K}$, which is simply the sample covariance
\begin{equation}
\hat{\boldsymbol{K}} = \frac{1}{N-1} \sum_j \boldsymbol{\eta}_j\boldsymbol{\eta}_j^{\rm T}.
\end{equation}
Alternatively, in the weak measurement case where $\sigma^2\gg \|K\|$, we can simply ignore $K$ by Taylor expanding Eq. \eqref{a:mle}:
\begin{equation}
\hat x = \frac{O_w(\boldsymbol{f})^{\rm T} \boldsymbol{r}}{\|O_w(\boldsymbol{f})\|^2} + O\left(\frac1{\sigma^2}\right).
\end{equation}
Dropping higher order terms, this estimator remains unbiased and has variance
\begin{align}
{\rm Var}[\hat x_{\rm MLE} ] &=  \frac{O_w(\boldsymbol{f})^{\rm T}(\boldsymbol{K} + \sigma^2  \boldsymbol{1})O_w(\boldsymbol{f})}{\|O_w(\boldsymbol{f})\|^4}\\
& =  \frac{\sigma^2}{\|O_w(\boldsymbol{f})\|^2} +\frac{O_w(\boldsymbol{f})^{\rm T}\boldsymbol{K} O_w(\boldsymbol{f})}{\|O_w(\boldsymbol{f})\|^4},
\end{align}
which is identical to the variance \eqref{a:var} of the exact maximum likelihood estimator.  Note that this illustrates \emph{exactly} how and why weak measurement (independent of any notion of post-selection or weak values) can aid in ``overcoming'' technical noise.

\section{FXS Hypothesis testing results}

\subsection{Hypothesis testing}

Let us consider the ``null hypothesis'' that no interaction is present: $x = 0$.  Standard statistical hypothesis testing would have us compute a ``test-statistic''.  In many cases, the most powerful is the \emph{likelihood ratio} test statistic \cite{Neyman1933Problem}:
\begin{equation}
D = -2 \log\left[\frac{\Pr(\boldsymbol{r},\boldsymbol{f}|x=0)}{\max_x \Pr(\boldsymbol{r},\boldsymbol{f}|x)}\right].
\end{equation}
For brevity, we define the symmetric matrix $\boldsymbol{Q} = (\boldsymbol{K} + \sigma^2  \boldsymbol{1})^{-1}$, so that 
 $\boldsymbol{r},\boldsymbol{f}|x \sim \mathcal N(x O_w(\boldsymbol{f}), \boldsymbol{Q}^{-1})$ and, in particular, $\boldsymbol{r},\boldsymbol{f}|x=0 \sim \mathcal N(0, \boldsymbol{Q}^{-1})$.
Then, the log-likelihood ratio becomes
\begin{equation}\label{test_stat}
D = \boldsymbol{r}^{\rm T} \boldsymbol{Q} \boldsymbol{r} - (\boldsymbol{r}- xO_w(\boldsymbol{f}))^{\rm T} \boldsymbol{Q}(\boldsymbol{r}- xO_w(\boldsymbol{f})).
\end{equation}

According to Wilk's theorem \cite{Wilks1938LargeSample}, under the null hypothesis the distribution of $D$ is asymptotically $\chi^2_N$: a $\chi$-squared distribution with $N$ degrees of freedom.  This fact, or a simple direct calculation analogous to one below, yields $\mathbb E_{\boldsymbol{r}|x=0}[D] = N$.  Now we compute the expected value of $D$ when an interaction is present.  This is a lengthy calculation but the result is conveniently simple.  Using  $\Pr(\boldsymbol{r}|\boldsymbol{f},x) \sim \mathcal N(x O_w(\boldsymbol{f}), \boldsymbol{Q}^{-1})$ an exercise in matrix algebra (see \cite[Sec. 6.2.2 Quadratic Forms]{Petersen2012Matrix})  reveals the first term in Eq. (\ref{test_stat}) is
\begin{align}
\mathbb E_{\boldsymbol{r}|\boldsymbol{f},x}[\boldsymbol{r}^{\rm T} \boldsymbol{Q} \boldsymbol{r}] &=\Tr[QQ^{-1}]+ x^2 O_w(\boldsymbol{f})^{\rm T} \boldsymbol{Q} O_w(\boldsymbol{f})= N + x^2 O_w(\boldsymbol{f})^{\rm T} \boldsymbol{Q} O_w(\boldsymbol{f}),
\end{align}
To calculate the second term we first note that the scalar $\hat x_{\rm MLE}$ is given by $\hat x_{\rm MLE} =  O_w(\boldsymbol{f})^T \boldsymbol{Q} \boldsymbol{r} / [ O_w(\boldsymbol{f})^T \boldsymbol{Q} O_w(\boldsymbol{f})  ] $ so that  $O_w(\boldsymbol{f}) \hat x_{\rm MLE} = O_w(\boldsymbol{f})O_w(\boldsymbol{f})^T \boldsymbol{Q} \boldsymbol{r} / [ O_w(\boldsymbol{f})^T \boldsymbol{Q} O_w(\boldsymbol{f})  ] $. Defining 
\begin{align}
  \boldsymbol{r}- xO_w(\boldsymbol{f}) =\underbrace{ \left ( \boldsymbol{1} - \frac{ O_w(\boldsymbol{f})  O_w(\boldsymbol{f})^T \boldsymbol{Q} }{O_w(\boldsymbol{f})^T \boldsymbol{Q}  O_w(\boldsymbol{f}) }\right)}_{\equiv \boldsymbol{A} } \boldsymbol{r}
\end{align}
we have
\begin{align}
 \mathbb E_{\boldsymbol{r}|\boldsymbol{f},x}[(\boldsymbol{r}- xO_w(\boldsymbol{f}))^{\rm T} \boldsymbol{Q}(\boldsymbol{r}- xO_w(\boldsymbol{f}))] & = \mathbb E_{\boldsymbol{r}|\boldsymbol{f},x}[\boldsymbol{r}^{\rm T}\boldsymbol{A}^{\rm T} \boldsymbol{Q}  \boldsymbol{A}  \boldsymbol{r} ]\\
 &=\underbrace{ \Tr{[\boldsymbol{A}^{\rm T} \boldsymbol{Q} \boldsymbol{A} \boldsymbol{Q}^{-1}   ]} }_{I}+ x^2 \underbrace{O_w(\boldsymbol{f})^{\rm T} \boldsymbol{A} ^{\rm T}  \boldsymbol{Q}   \boldsymbol{A}O_w(\boldsymbol{f})}_{II}.
 \end{align}
Simplifying the sub-expression $I$ gives
\begin{align}
I &=\Tr \left [  \boldsymbol{1}
- \frac{  \boldsymbol{Q}O_w(\boldsymbol{f})O_w(\boldsymbol{f})^{\rm T}    }{ O_w(\boldsymbol{f})^{\rm T} \boldsymbol{Q} O_w(\boldsymbol{f})  }
- \frac{  \boldsymbol{Q}O_w(\boldsymbol{f})O_w(\boldsymbol{f})^{\rm T}    }{ O_w(\boldsymbol{f})^{\rm T} \boldsymbol{Q} O_w(\boldsymbol{f})  }
 + \frac {  \boldsymbol{Q}O_w(\boldsymbol{f})O_w(\boldsymbol{f})^{\rm T} \boldsymbol{Q}  O_w(\boldsymbol{f})O_w(\boldsymbol{f})^{\rm T}   } {  (O_w(\boldsymbol{f})^{\rm T} \boldsymbol{Q} O_w(\boldsymbol{f}) )^2  }
  \right ] \\
 &  = 0
\end{align}
where we have used the cyclic property of the trace. To simplify $II$ we first define 
\begin{align}
  \boldsymbol{A} = \Bigg ( \boldsymbol{1} - \underbrace{\frac{ O_w(\boldsymbol{f})  O_w(\boldsymbol{f})^T \boldsymbol{Q} }{O_w(\boldsymbol{f})^T \boldsymbol{Q}  O_w(\boldsymbol{f}) }    }_{\equiv \boldsymbol{B} }  \Bigg)
\end{align}
then we have
\begin{align}
II 
&=O_w(\boldsymbol{f})^{\rm T} [\boldsymbol{1} - \boldsymbol{B}^{\rm T}]  \boldsymbol{Q} [\boldsymbol{1} - \boldsymbol{B}] O_w(\boldsymbol{f}) \\
&= O_w(\boldsymbol{f})^{\rm T}\boldsymbol{Q}O_w(\boldsymbol{f}) - O_w(\boldsymbol{f})^{\rm T}\boldsymbol{B}^{\rm T} \boldsymbol{Q} O_w(\boldsymbol{f})- O_w(\boldsymbol{f})^{\rm T}\boldsymbol{Q}\boldsymbol{B}O_w(\boldsymbol{f}) + O_w(\boldsymbol{f})^{\rm T}\boldsymbol{B}^{\rm T} \boldsymbol{Q}\boldsymbol{B}O_w(\boldsymbol{f})   \\
&= \Bigg\{  O_w(\boldsymbol{f})^{\rm T}   \boldsymbol{Q}    O_w(\boldsymbol{f})
- \frac{  O_w(\boldsymbol{f})^{\rm T}\boldsymbol{Q}  O_w(\boldsymbol{f})O_w(\boldsymbol{f})^{\rm T}\boldsymbol{Q} O_w(\boldsymbol{f}) }{ O_w(\boldsymbol{f})^{\rm T} \boldsymbol{Q} O_w(\boldsymbol{f})  }
- \frac{  O_w(\boldsymbol{f})^{\rm T}\boldsymbol{Q}  O_w(\boldsymbol{f})O_w(\boldsymbol{f})^{\rm T}\boldsymbol{Q} O_w(\boldsymbol{f}) }{ O_w(\boldsymbol{f})^{\rm T} \boldsymbol{Q} O_w(\boldsymbol{f})  }\nonumber\\
&\quad\quad
+\frac{ O_w(\boldsymbol{f})^{\rm T}\boldsymbol{Q} O_w(\boldsymbol{f})O_w(\boldsymbol{f})^{\rm T}\boldsymbol{Q}  O_w(\boldsymbol{f})O_w(\boldsymbol{f})^{\rm T}\boldsymbol{Q}O_w(\boldsymbol{f})  }{ (O_w(\boldsymbol{f})^{\rm T} \boldsymbol{Q} O_w(\boldsymbol{f}) )^2 
}   \Bigg\}\\
 &  = O_w(\boldsymbol{f})^{\rm T}   \boldsymbol{Q}    O_w(\boldsymbol{f}) - O_w(\boldsymbol{f})^{\rm T}\boldsymbol{Q}  O_w(\boldsymbol{f})- O_w(\boldsymbol{f})^{\rm T}\boldsymbol{Q}  O_w(\boldsymbol{f})+O_w(\boldsymbol{f})^{\rm T}   \boldsymbol{Q}    O_w(\boldsymbol{f})\\
 &=0.
\end{align}
Thus, we have $\mathbb E_{\boldsymbol{r}|\boldsymbol{f},x}[D]=
\mathbb E_{\boldsymbol{r}|\boldsymbol{f},x}[\boldsymbol{r}^{\rm T} \boldsymbol{Q} \boldsymbol{r}] - \mathbb E_{\boldsymbol{r}|\boldsymbol{f},x}[(\boldsymbol{r}- xO_w(\boldsymbol{f}))^{\rm T} \boldsymbol{Q}(\boldsymbol{r}- xO_w(\boldsymbol{f}))] $ so that
\begin{equation}
\mathbb E_{\boldsymbol{r}|\boldsymbol{f},x}[D] = N + x^2 O_w(\boldsymbol{f})^{\rm T} \boldsymbol{Q} O_w(\boldsymbol{f}).
\end{equation}
We have already encountered the term $O_w(\boldsymbol{f})^{\rm T} \boldsymbol{Q} O_w(\boldsymbol{f}) = O_w(\boldsymbol{f})^{\rm T}(\boldsymbol{K} + \sigma^2\boldsymbol{1})^{-1} O_w(\boldsymbol{f})$ above.  The Taylor expansion shows
\begin{equation}
\mathbb E_{\boldsymbol{r}|\boldsymbol{f},x}[D] = N + \frac{x^2}{\sigma^2} \|O_w(\boldsymbol{f})\|^2 + O\left(\frac1{\sigma^4}\right).
\end{equation}
Recalling the mean of $\|O_w(\boldsymbol{f})\|^2$ is
\begin{equation}
\mathbb E_{\boldsymbol{f}}[\|O_w(\boldsymbol{f})\|^2] = N\langle i|O^2|i\rangle,
\end{equation}
we see that, ignoring higher order terms,
\begin{equation}
\mathbb E_{\boldsymbol{r}, \boldsymbol{f}|x}[D] = N\left(1  + \frac{x^2 \langle i|O^2|i\rangle}{\sigma^2}\right).
\end{equation}

Thus, if an interaction of strength $x$ is present, it will on average exceed our expectation of $D$ by a factor of
\begin{equation}
1  + \frac{x^2 \langle i|O^2|i\rangle}{\sigma^2},
\end{equation}
and we again see that the weak value amplification technique is suboptimal.

\end{document}